\documentclass[%
prl,twocolumn,
superscriptaddress,
 amsmath,amssymb,
 aps,
]{revtex4-1}

\usepackage{graphicx}
\usepackage{dcolumn}
\usepackage{bm}
\usepackage{xcolor}
\usepackage{mathtools}
\usepackage{bbold}

\newcommand{\ndg}{{\phantom{\dagger}}}
\newcommand{\dg}{\dagger}
\newcommand{\ew}[1]{\left\langle #1 \right\rangle}
\newcommand{\ket}[1]{| #1 \rangle}
\newcommand{\bra}[1]{\langle #1 |}

\newcommand{\ketbra}[2]{\left|#1\right\rangle\hspace{-1.1mm}\left\langle #2 \right|}
\newcommand{\ketbras}[2]{\left|#1\right\rangle_{SS}\hspace{-0.9mm}\left\langle #2 \right|}
\newcommand{\ketbrar}[2]{\left|#1\right\rangle_{RR}\hspace{-0.9mm}\left\langle #2 \right|}

\newcommand{\td}{\text{d}}
\newcommand{\avg}[1]{\langle\hspace{-0.5mm}\langle #1 \rangle\hspace{-0.5mm}\rangle}

\begin{document}

\preprint{APS/123-QED}

\title{A quantum optical realization of the Ornstein-Uhlenbeck process via simultaneous action of white noise and feedback}

\author{Alexander Carmele}
\email{alex@itp.tu-berlin.de}
\affiliation{Department of Physics, University of Auckland, Private Bag 92019, Auckland, New Zealand}
\affiliation{Technische Universit\"at Berlin, Institut f\"ur Theoretische Physik, Nichtlineare Optik und Quantenelektronik, Hardenbergstra{\ss}e 36, 10623 Berlin, Germany}
\author{Scott Parkins}
\affiliation{Department of Physics, University of Auckland, Private Bag 92019, Auckland, New Zealand}
\affiliation{The Dodd-Walls Centre for Photonic and Quantum Technologies, New Zealand}
\author{Andreas Knorr}
\affiliation{Technische Universit\"at Berlin, Institut f\"ur Theoretische Physik, Nichtlineare Optik und Quantenelektronik, Hardenbergstra{\ss}e 36, 10623 Berlin, Germany}

\date{\today}

\begin{abstract}
We establish an important connection between coherent quantum feedback and the Ornstein-Uhlenbeck process in quantum optics.
We show that an emitter with fluctuating energy levels in front of a mirror results in an Ornstein-Uhlenbeck process for electronic populations, although the fluctuation of the energy levels is assumed to be uncorrelated in time and space.
Based on a Heisenberg equation of motion description of the quantum feedback dynamics, we discuss additionally the impact of phase noise on the population dynamics and provide examples in which noise itself is not detrimental but supports and enhances typical features of quantum  feedback such as self-stabilization.
\end{abstract}

\maketitle

The fluctuation-dissipation theorem relates a fluctuating force to its corresponding imposed friction and expresses in this sense the balance between noise and dissipation \cite{kubo1966fluctuation,zwanzig2001nonequilibrium,gardiner2004quantum,gardiner2009stochastic,risken}.
It is widely used in non-equilibrium statistical mechanics, in which correlation functions replace the partition function as the crucial quantity to characterize properties and the long-time behavior of the system under study
\cite{nelson1967dynamical,kubo2012statistical,sekimoto}.
Such non-equilibrium systems are often described by 
the Ornstein-Uhlenbeck (O-U) process  \cite{uhlenbeck1930theory,wax1954selected,wang1945theory}
which formalizes and generalizes Einstein's description of Brownian motion with 
\begin{align}
\dot u = -\gamma u + F_t,
\end{align}
where $u$ is the particle velocity, $\gamma$ the friction and $F_t$ the corresponding fluctuating force, which enforces thermal equilibrium in the long-time limit \cite{einstein1905motion}.
Based on the O-U process which considers a noise correlation $\avg{F_t F_s}=\Gamma \exp[-\gamma |t-s|]$, a mean-squared displacement of a particle can be derived and, using the Einstein-Smoluchowski relation \cite{einstein1905motion,smoluchowski1918versuch,philipse2018}, reads:
\begin{align} \label{eq:mean_square_displacement}
\avg{[x(t)]^2}=\frac{A_0}{\gamma^2}    
\left[
\gamma t+e^{-\gamma t} -1 
\right],
\end{align}
with $A_0$ being a constant fulfilling the corresponding self-diffusion coefficient \cite{uhlenbeck1930theory,wang1945theory}.
These formulas are applicable to a wide range of physical systems beyond the particular case of Brownian motion and enable the determination of viscosity \cite{huang2011direct}, thermal and electrical conductivity  \cite{shoghl2016electrical,saffman1975brownian}.
Examples include light scattered by diffusing molecules in spectroscopy measurements \cite{pusey1978intensity,maret1987multiple,burada2009diffusion}
and the derivation of absorption coefficients such as the Lambert-Beer law determined by the electric-dipole moments of the substance
\cite{zwanzig2001nonequilibrium,risken,philipse2018,ishii2005single}.
In this study, we will derive the mean-square displacement formula in Eq.~\eqref{eq:mean_square_displacement} for electronic populations of an atom excited by quantum optical fields \cite{gardiner2004quantum,gardiner2015quantum}:
Describing the radiative decay of an excited atom in the vicinity of a mirror \cite{hoi2015probing,svidzinsky2018excitation,hetet2011single,wilson2003vacuum,glaetzle2010single,dubin2007photon}, the coherent quantum feedback dynamics lead straightforwardly to the O-U process if fluctuations of the energy levels are accounted for during the emission dynamics.
Interestingly, this result can be attributed to a mirror charge-induced dipole-dipole correlation due to the mirror, as has been derived by Zwanzig \cite{zwanzig2001nonequilibrium}, only here retarded in time and of purely single quantum nature.
The main ingredient for describing spontaneous emission used here beyond the Wigner-Weisskopf limit \cite{weisskopf1997naturliche}, in which the vacuum field amplitude is considered constant with respect to the frequency of the emitted photon, is the coherent quantum feedback mechanism \cite{dorner-half-cavities,Cook86feedback,milonni,cook1987quantum,cook2018input,nemet2016enhanced,faulstich2018unraveling,pichler2015photonic,crowder2019quantum,barkemeyer2019revisiting}.
This kind of feedback has been proven to be a versatile strategy to steer and control systems non-invasively, and is not related to measurement-based, or invasive quantum feedback control \cite{ChaosControl2008,sfb910}.
The coherent and non-Markovian nature imposes  quantum interferences between present and past system states onto the dynamics and allows for interesting two-photon processes \cite{droenner2019quantum,pichler2015photonic}, enhanced entanglement and non-classical photon statistics \cite{lu2017intensified}, dimerization \cite{guimond2017delayed,guimond2016chiral}, and a stabilization of quantum coherence due to interference effects between incoming and outgoing probability waves \cite{kabuss2015analytical}. 
A typical paradigm for such processes is the formation of dark states and subsequently emerging population trapping \cite{multimode,carmele2019retardation}.
In the following, we will establish an important connection between the non-Markovian quantum feedback process and the generation of an Ornstein-Uhlenbeck-type of noise correlations.
For this, we derive an equation of motion via the Heisenberg picture for the microscopic coherence operator subjected to white noise fluctuations and quantum feedback.
Importantly, we show that white noise-based energy fluctuations are not necessarily detrimental to quantum feedback effects, as it counteracts destructive and unwanted interference effects between the incoming and outgoing photon emission processes, and even supports
population trapping for certain quantum feedback phase relations.
%

{\it Model.}
In this section, we discuss the impact of white
noise on the emission dynamics of a two-level system in front of a mirror.
The Hamiltonian of the system reads $(\hbar=1)$:
\begin{align}
\label{eq:Hamiltonian}
H = (\omega_0+F_t) P^\dg P\hspace{-0.1cm} + \hspace{-0.2cm} \int \hspace{-0.1cm}\td\omega \hspace{-0.1cm} \left[
r^\dg_\omega
\left( \frac{\omega r^\ndg_\omega}{2} + g^*_\omega P \right)\hspace{-0.1cm}+ \text{h.a.} \right]
\end{align}
where $ P=\ketbra{g}{e}$ denotes the microscopic coherence operator from the excited- 
$ \ket{e} $ to the ground-state $ \ket{g}$ of the two-level system with a transition energy of $\hbar\omega_0$
\cite{dorner-half-cavities,Cook86feedback,milonni,cook1987quantum,cook2018input,nemet2016enhanced,faulstich2018unraveling,pichler2015photonic,crowder2019quantum,barkemeyer2019revisiting,kabuss2016}.
The radiative continuum is included via the photon creation and annihilation operators $ r^{(\dg)}_\omega $ for a photon in the mode $ \omega = c k$ ($c$: the speed of light in the waveguide) with bosonic commutation relations: $[r^{\ndg}_\omega ,r^{\dg}_{\omega'}]=\delta(\omega-\omega')$.
The coupling between the emitter and the radiative continuum is denoted by $ g_\omega=g_0 \sin(\omega\tau/2) $ and includes the mirror imposed boundary condition at a distance $L$ between mirror and atom with a strength of $g_0$.
The length defines the feedback round trip time with $ \tau=2L/c $.
$F_t$ describes a stochastic force acting upon the excited level of the two-level system and models, e.g. a spectral diffusion process \cite{gardiner2009stochastic,hanggi2005fundamental,vural2018two}.
We assume throughout our analysis a Gaussian white noise with vanishing average $\avg{F_t}=0$ and $\delta$-correlated correlation function $\avg{F_t F_s}=\gamma\delta(t-s)$.
Next, we  solve this model in the Heisenberg picture \cite{kira2011semiconductor}.
The equation of the Heisenberg operator $P^\dg(t)=U^\dg(t)P^\dg U(t)$ with $U(t)=\exp\left[-iH t \right]$ using $\dot P^\dg(t)=i[H,P^\dg(t)]$ reads
in the rotating-frame
\begin{align}
\dot P^\dg(t) 
\hspace{-0.1cm}
=
\hspace{-0.1cm}
iF_t P^\dg(t)
\hspace{-0.1cm}
+
\hspace{-0.1cm}
i\hspace{-0.1cm}\int\hspace{-0.1cm} \td\omega 
g^*_\omega e^{i(\omega-\omega_0)t}
r^\dg_\omega(t)[P(t),P^\dg(t)] .
\end{align}
The coherence operator couples to the inversion and to the quantized light field.
Starting with an initial condition at $t=0$, the goal is to solve for the quantized light-field exactly by integrating out the equation of motion of the photon creation operator:
\begin{align}
r^\dg_\omega(t)
=
r^\dg_\omega(0)
+
ig_\omega
\int_0^t \td t_1
e^{-i(\omega-\omega_0)t_1}
P^\dg(t_1).
\end{align}
This equation allows us to write down the Heisenberg-Langevin equation of motion. Within the one-electron assumption for the two-level system, the inversion operator can be written as $[P(t),P^\dg(t)]=1-2P^\dg(t)P(t)$
and for the dynamics of the coherence operator follows
\begin{align}\notag
\dot P^\dg(t)  
=&-[\Gamma-iF_t] P^\dg(t)
+\Gamma e^{-i\omega_0\tau}
P^\dg(t-\tau)\theta(t-\tau) \\ \notag
&-2 \Gamma e^{-i\omega_0\tau}
P^\dg(t-\tau) 
P^\dg(t)P(t)
\theta(t-\tau) \\ 
&+ig_0 R^\dg(t)[P(t),P^\dg(t)]
\label{eq:full_pdg},
\end{align}
where $R^\dg(t)=\int d\omega  r^\dg_\omega(0) \sin(\omega\tau/2)\exp[i(\omega-\omega_0)t]$ includes the quantum noise contribution to conserve the commutation for all times with $\Gamma=g_0^2\pi/2$.
Clearly, the signal $P$ at the feedback delay time $\tau$ occurs in Eq.~\eqref{eq:full_pdg}.
In the following, we show that the second and last (third) line vanishes in the case of a reservoir initially in the vacuum state
and a system described by the Hamiltonian in Eq.~\eqref{eq:Hamiltonian}.
The solution of Eq.~\eqref{eq:full_pdg} is derived for every $\tau$-interval iteratively \cite{dorner-half-cavities,kabuss2015analytical,kabuss2016}.
For the time interval, $t\in[0,\tau]$,
we evaluate the matrix element of the coherence operator $P^*_{ij}(t)=\bra{i,\text{vac}}P^\dg(t)\ket{j,\text{vac}}$
with $\ket{j,\text{vac}}=\ket{j}_S\ket{\text{vac}}_R$ and $j$ either $e$ or $g$ for the system state and the reservoir in the vacuum state.
The dynamics of the polarization reduces to:
\begin{align}
\dot P^*_{ij}(t)  
=&-[\Gamma-iF_t] P^*_{ij}(t), \\
\label{eq:zero_interval}
P^*_{ij}(t)  =&
e^{-\Gamma t+i\phi(t,0)} P^*_{ij}(0),
\end{align}
contributing only for $i=e$ and $j=g$
and $\phi(b,a):=\int_a^b F_{t'} dt'$.
Note that the matrix element does not represent the expectation value.
However, the expectation value can be fully expressed by its corresponding matrix elements, e.g. $\ew{P^\dg(t)P(t)}=|P^*_{eg}(t)|^2$.
For the second interval, $t\in[\tau,2\tau]$,
the dynamics of the matrix element reads:
\begin{align}\label{eq:first_interval}
\dot P^*_{ij}(t)  
=&-\Gamma P^*_{ij}(t)
+\Gamma e^{-i\omega_0\tau}
P^*_{ij}(t-\tau)\\ \notag
&-2 \Gamma e^{-i\omega_0\tau}
\bra{i,\text{vac}}
P^\dg(t-\tau) 
P^\dg(t)P(t)
\ket{j,\text{vac}} .
\end{align}
Due to the occuring time delay, we can use $P_{ij}$
for $i=e$ and $j=g$ from Eq.~\eqref{eq:zero_interval} in
Eq.~\eqref{eq:first_interval} to evaluate the second line.
For this, we insert now the unity relation $\mathbb{1}=\sum_{i=e,g}
\ketbras{i}{i}\otimes\left(\ketbrar{\text{vac}}{\text{vac}}+\int d\omega \ketbrar{1_\omega}{1_\omega} \right)$ to evaluate the 
correlation between the "time-nonlocal" microscopic coherence and the time-local population density, and taking into account that only $\bra{i,\text{vac}} P^\dg(t-\tau)\ket{g,\text{vac}}$ can contribute non-trivially:
\begin{align}  
&\bra{i,\text{vac}}
P^\dg(t-\tau) 
P^\dg(t)P(t)
\ket{j,\text{vac}}
\\ \notag
&=\bra{i,\text{vac}}
P^\dg(t-\tau) 
\ket{g,\text{vac}}
\bra{g,\text{vac}}
P^\dg(t)P(t)
\ket{j,\text{vac}},
\end{align}
having reduced the problem to the
matrix element $\bra{g,\text{vac}}
P^\dg(t)P(t)
\ket{j,\text{vac}}$.
If we now again insert a unity operator between the operators $P^\dg(t)P(t)$,
we reduce this quantity again
into further products of matrix elements.
Since we know, that only $P^*_{eg}(t)$ 
contributes initially in the first time interval,
the quantity vanishes identically in the case of the Hamiltonian dynamics in Eq.~\eqref{eq:Hamiltonian}
due to $\bra{g,\text{vac}} P^\dg(t)\ket{\phi}=0$ for arbitrary $\ket{\phi}$, and therefore we can conclude that, assuming an initially empty reservoir, the matrix elements of the microscopic coherence operator is governed by the dynamics for all times $t$: %
\begin{align}\notag
\dot P^*_{eg}(t)  
=&(iF_t-\Gamma) P^*_{eg}(t)
+\Gamma e^{-i\omega_0\tau}
P^*_{eg}(t-\tau)\theta(t-\tau).
\end{align}
In the case of $F_t\equiv0$, this equation can be solved in the Laplace domain \cite{dorner-half-cavities,glaetzle2010single,sinha2019non,kabuss2015analytical,crowder2019quantum}, yielding the following known dynamics valid for all $t$:
\begin{align}
\label{eq:coherence_solution_wo_noise}
P^*_{eg}(t)\hspace{-0.1cm}=\hspace{-0.2cm} \sum_{n=0}^\infty \frac{e^{-\Gamma t}}{n!}
\hspace{-0.1cm}
\left[ 
\Gamma
e^{-i\omega_0\tau + \Gamma \tau} (t-n \tau)
\right]^n \Theta(t-n \tau).
\end{align}
However, in the following we are interested in the
case of a non-vanishing noise $F_t\neq0$, and
the equation can only be solved via subsequent integration with respect to time. 
For example, for $t\in[0,3\tau]$:
\begin{align}
P^*_{eg}(t)  \notag
=
&e^{-\Gamma t+i\phi(t,0)} \bigg[\theta(t)
+\theta(t-\tau)\Gamma e^{-i\omega_0\tau+\Gamma\tau}
N(t,\tau) \\ 
&+\theta(t-2\tau)
(\Gamma e^{-i\omega_0\tau+\Gamma\tau})^2 M(t,2\tau) 
\bigg]\label{eq:p_eg_eom}
\end{align}
with $\phi(b,a)=\int_a^b F_{t'} dt'$ and the definitions
\begin{align*}
N(t,\tau) 
&:=
\int_\tau^t \td t_1
e^{-i\phi(t_1,t_1-\tau)} \\
M(t,2\tau) &:=
\int_{2\tau}^t \td t_1
e^{-i\phi(t_1,t_1-\tau)}
\int_{\tau}^{t_1-\tau} \hspace{-0.6cm} \td t_2
e^{-i\phi(t_2,t_2-\tau)},
\end{align*}
which recover in the limit of $\gamma~\rightarrow~0$, the solution given in Eq.~\eqref{eq:coherence_solution_wo_noise}, i.e. $N(t,\tau)=(t-\tau)$ and $M(t,2\tau)=(t-2\tau)^2/2 $ \footnote{For details see the supplemental material.}.
In the following, we are interested in the cases 
(i) $0\le t\le2\tau$ and (ii) $0\le t\le3\tau$.
For (i), we show that the quantum feedback contribution leads to a Ornstein-Uhlenbeck process for the population dynamics due to the assumed white noise correlation.
In (ii), we discuss the impact of phase noise on the coherent quantum feedback dynamics and show that it need not necessarily be detrimental to quantum feedback effects, as disadvantageous destructive interferences, if they occur, are suppressed.
{\it (i) Ornstein-Uhlenbeck}.
Due to the white noise contribution, we cannot use 
the solution derived via the Lambert W-function in Eq.~\eqref{eq:coherence_solution_wo_noise}.
We use Eq.~\eqref{eq:p_eg_eom} to evaluate for 
the population dynamics.
The solution of the population dynamics reads for $\tau \le t\le2\tau$:
\begin{align}
|P^*_{eg}(t)|^2   
&= \ew{P^\dg(t)P(t)}=
e^{-2\Gamma t}  \\ \notag
&
+
2\Gamma e^{-\Gamma(2t-\tau)}
\int_0^{t-\tau} \hspace{-0.5cm} \td s'
\text{Re}
\left[
e^{i\omega_0\tau}
e^{i\phi(s'+\tau,s')}
\right] 
\\ \notag
&
+
\left(\Gamma e^{-\Gamma (t-\tau)}\right)^2
\int_0^{t-\tau}\hspace{-0.5cm} \td s'
\int_0^{t-\tau}\hspace{-0.5cm} \td s
e^{-i\phi(s+\tau,s)
+i\phi(s'+\tau,s')} 
\end{align}
where we used the property: 
$\phi(s+\tau,0)-\phi(s,0)
=\phi(s+\tau,s)$ \footnote{See supplemental material for the detailed derivation.}.
Evaluating the noise correlation, 
we specify the result to $\omega_0\tau/(2\pi)=n$ for $n$ integer:
\begin{align}
\label{eq:solution_2tau}
\avg{|P^*_{eg}(t)|^2}   
=
&e^{-2\Gamma t} \\ \notag
&\bigg[
1 +
\frac{2\Gamma^2}{\gamma^2} 
e^{2\Gamma\tau}  \left( \gamma(t-\tau)
+
e^{-\gamma(t-\tau)}
-1
\right)
\bigg],
\end{align}
which is the main result of our analysis.
For non-vanishing noise $F_t\neq0$, the initial, non-convoluted white noise contribution in the emission process takes the form of an Ornstein-Uhlenbeck process, as can be seen from Eq.~\eqref{eq:solution_2tau}.
This is a remarkable result: In the first-$\tau$ interval the noise averaged population dynamics is  a superposition of a radiative decay with an O-U process:
\begin{align} \notag
\avg{|P^*_{eg}(t)|^2}   
&=
e^{-2\Gamma t}
\left( 1+
\int_0^{t-\tau} \hspace{-0.6cm} \td s
\int_0^{t-\tau} \hspace{-0.6cm} \td s^\prime
\avg{F(s)F(s^\prime)}|_{U-O} 
\right),
\end{align}
with the noise correlation 
\begin{align}\notag
\avg{F(t_1)F(t_2)}|_{U-O}=\Gamma_\tau e^{-\gamma|t_1-t_2|}
= \Gamma^2 e^{2\Gamma\tau-\gamma|t_1-t_2|}.
\end{align}
Herewith, we have accomplished our goal to re-derive Eq.~\eqref{eq:mean_square_displacement} in quantum optics.
In fact, the white noise contribution is
transformed via a feedback mechanism into an O-U noise process.
Excitingly, this opens  completely new interpretation
horizons concerning the O-U-process.
In terms of metrology, the quantum feedback mechanism  effectively acts as low-pass filter to convolute the initial white noise into an O-U process \cite{bibbona2008ornstein,risken}.
This has interesting implications for the interpretation and application of the U-O process and its role in non-equilibrium statistical mechanics in general \cite{sekimoto,zwanzig2001nonequilibrium,philipse2018} and quantum optics in particular \cite{gardiner2009stochastic,gardiner2004quantum,gardiner2015quantum,reiter2019distinctive,carmele2019non,luker2019review}, if non-Markovianity and irreversibility are equally assumed \cite{ChaosControl2008,sfb910,gardiner2004quantum,loos2019heat,loos2019non,loos2017force}.
For completeness, we note that in the limit of $\gamma \rightarrow 0$, 
the exponential, $(\gamma(t-\tau) +\exp[-\gamma(t-\tau)]-1)\rightarrow \gamma^2(t-\tau)^2/2$ is expanded and recovers the analytical solution without noise, cf.~Eq.~\eqref{eq:coherence_solution_wo_noise}.

\begin{figure}[t!]
\includegraphics[width=1.0\linewidth]{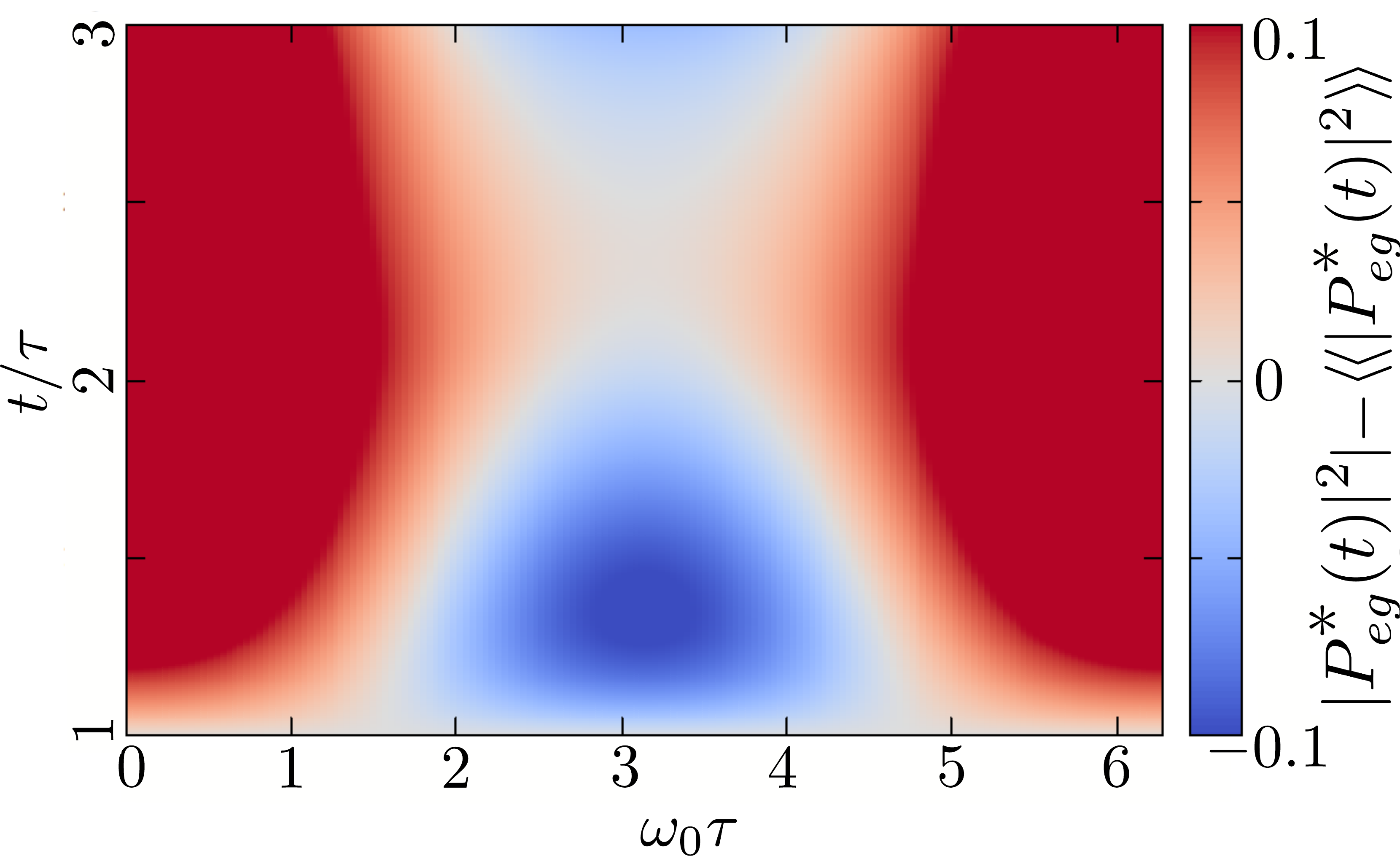}
\caption{The difference between the population dynamics in the interval $t/\tau\in[1,3]$ with $\avg{|P^*_{eg}(t)|^2}$ and without dephasing $|P^*_{eg}(t)|^2$ for different phase choices $\varphi=\omega_0\tau$. Pure dephasing is for population trapping advantageous if the phase value is in the interval $[\pi/2,3\pi/2]$.}
\label{fig:heatmap}
\end{figure}

{\it (ii) Suppression of destructive interferences.}
The Heisenberg equation of motion formulation of coherent quantum feedback allows to investigate the role of phase noise in typical radiative decay processes in the presence of delay.
After having demonstrated that white noise fluctuations are transformed into an O-U process, we  now study the impact of these fluctuations on the population dynamics up to $3\tau.$
For the population dynamics, we find the following expression for ($2\tau\le t\le3\tau$):
\begin{widetext}
\begin{align}\notag
\avg{|P^*_{eg}(t)|^2} 
= e^{-2\Gamma t}
\bigg[ 
1
&+2\Gamma e^{\Gamma\tau}\cos(\omega_0\tau) 
\avg{N(t,\tau)} 
+\Gamma^2 e^{2\Gamma\tau} 
\avg{N(t,\tau)N^*(t,\tau)}  +2\Gamma^2 e^{2\Gamma\tau}\cos(2\omega_0\tau) 
\avg{M(t,2\tau)} \\ 
&+2\Gamma^3 e^{3\Gamma\tau}\cos(\omega_0\tau) 
\avg{N^*(t,\tau)M(t,2\tau)} 
+\Gamma^4 e^{4\Gamma\tau} 
\avg{M(t,2\tau)M^*(t,2\tau)}  \bigg],
\label{eq:solution_3tau_w_dephasing}
\end{align}
\end{widetext}
with $M(t,2\tau)$ and $N(t,\tau)$ defined after Eq.~\eqref{eq:p_eg_eom}.
Evaluating the noise integrals involves up to four time-ordered integrals and its corresponding  noise-noise correlations. This yields lengthy expressions which are given explicitly in the supplemental material.
\begin{figure}[t!]
\includegraphics[width=0.8\linewidth]{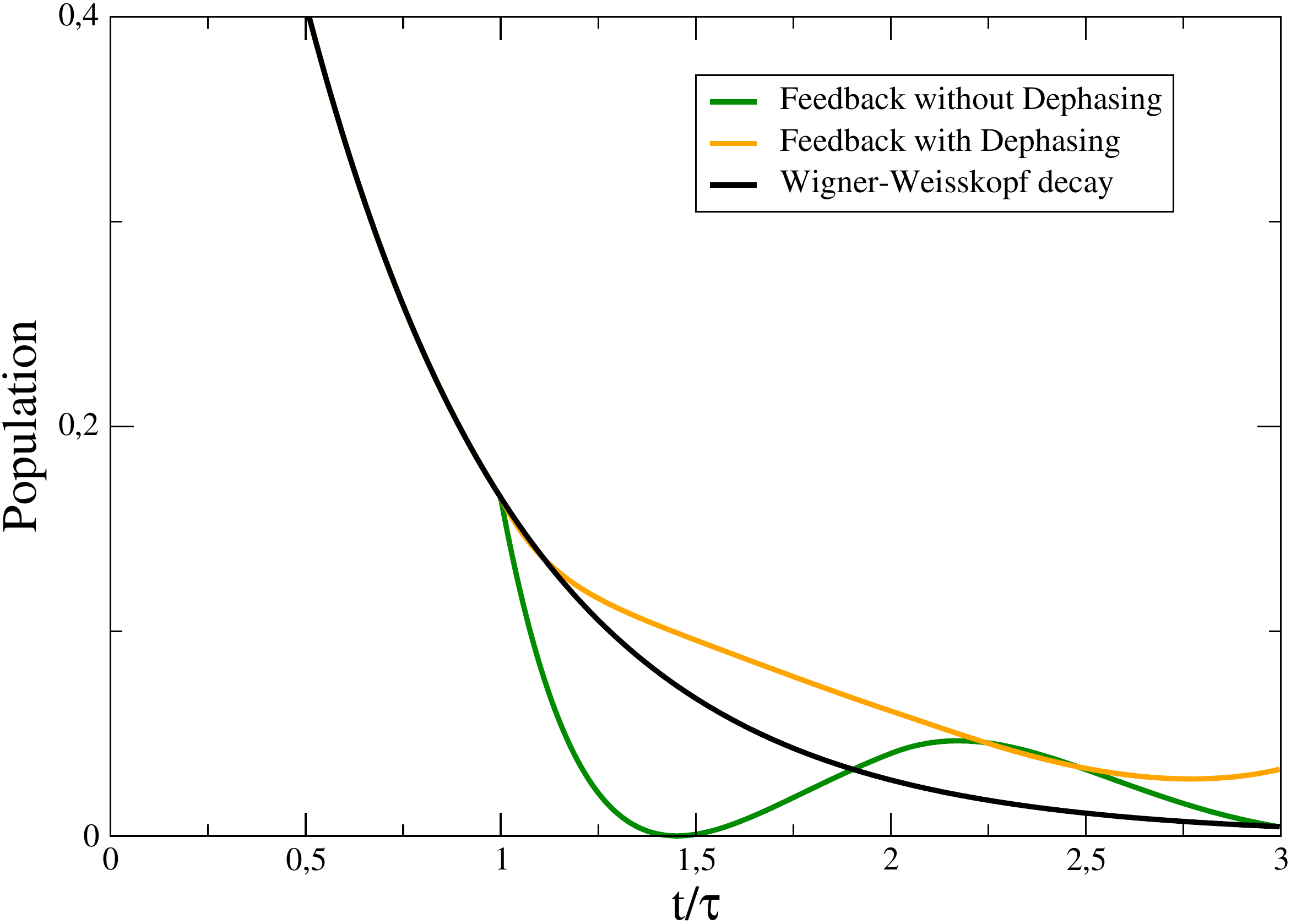}
\caption{The population dynamics in the interval $t/\tau\in[0,3]$ and for the phase $\varphi=\omega_0\tau=3.3$ in case of the Wigner-Weisskopf decay without feedback (black line), and with feedback but without dephasing (green line), and with feedback and dephasing (orange line). Interestingly, pure dephasing suppresses detrimental oscillations occurring due to the phase choice and remains well-above the Wigner-Weisskopf case.}
\label{fig:pop_vs_time}
\end{figure}
Often, the goal of quantum feedback is to stabilize electronic populations due to destructive interference between absorption and
re-emission events \cite{pichler2015photonic,kabuss2015analytical}.
This leads in the perfect case of $\varphi=\omega_0\tau/(2\pi)=n$ with n integer to population trapping or a bound state in the continuum
\cite{multimode,PhysRevLett.122.073601}. 
For any phase unequal to a multiple of $\varphi\neq\omega_0\tau/(2\pi)=n$ and without phase noise $F_t\equiv0$, the population decays inevitably.
We show now that the population decay can be slowed down in the presence of phase noise despite a disadvantageous phase choice. 
In Fig.~\ref{fig:pop_vs_time}, the dynamics of the population is depicted for the case without feedback (Wigner-Weisskopf case, black line), without noise but with feedback (green line), and with feedback and with noise (orange line) for a phase of $\varphi=\omega_0\tau=3.3$.
As can be seen, phase noise helps to suppress the destructively interfering parts of the solution in Eq.~\eqref{eq:solution_3tau_w_dephasing} proportional to $\cos(n\omega_0\tau)$ with n integer.
These contributions enforce damped oscillations of the population, leading eventually to a complete decay of the electronic excitation into the reservoir with zero excitation left in the emitter (green line).
However, these contributions are strongly affected by the phase noise.
Here, in the transient regime, noise helps to slow down the decay of the electronic population and prevents it from decaying rapidly to zero (orange line). 
In Fig.~\ref{fig:pop_vs_time}, the population in the emitter in case of finite $F_t$ is larger or for a short time ($t/\tau \approx 2.4$) equal/slightly less compared to the case with vanishing phase noise.
As a comparison, we plot the dynamics imposed by just the Wigner-Weisskopf case (black line).
The case with dephasing is always larger than the Wigner-Weisskopf dynamics, whereas the case with feedback and no phase noise oscillates following the decay of the Wigner-Weisskopf solution, indicating an inevitable complete decay for undisturbed feedback.
Interestingly, the solution with feedback shows a non-monotonous behavior to the end of the third $\tau$-interval where population is gained.
These results, however, depend on the choice of the feedback phase
$\varphi=\omega_0\tau$.
This indicates that the choice of the delay time provides another control parameter to optimize the phase noise action on the population number:
In Fig.~\ref{fig:heatmap}, we plot the difference between the population with and without noise: 
$|P^*_{eg}(t)|^2 - \avg{|P^*_{eg}(t)|^2}$.
We clearly see that phase noise is detrimental to population trapping in the vicinity of the phase $\varphi=0$, as this phase choice renders the destructive interference terms already unimportant, i.e. for $\omega_0\tau\in(0,\pi/2)$ and $\omega_0\tau\in(3\pi/2,2\pi)$.
However, for phases in between $(\pi/2,3\pi/2)$, the
population is enhanced due to noise.
We conclude that for $\varphi\neq2\pi$, phase noise still allows  for important feedback effects relying on the population trapping mechanism.
We like to add that the calculation of the noise contribution up to $3\tau$ is straightforward but already lengthy in this approach.
So far, the long-time limit is not accessible, as for every $\tau$-interval, the noise contributions need to be evaluated separately due to the time-reordering.
Here, a fully quantum mechanical model is envisioned to also investigate the long-time behavior without evaluating every time-interval individually.
{\it Conclusion.} We have studied the impact of white noise on the radiative decay dynamics of an atom in front of a mirror. 
We have shown that the white noise contribution, here a fluctuation of the excited energy level, leads to an Ornstein-Uhlenbeck process.
Herewith, we establish an interesting relation between non-Markovian feedback processes and the fluctuation-dissipation theorem and allow for interpretations of the O-U processes in non-equilibrium statistical mechanics in terms of non-Markovianity and feedback acting as a low-pass filtered white noise. 
Furthermore, we discussed the impact of phase  noise on the population trapping dynamics and show advantageous features in a wide range of phase choices.
\ \\
AC and AK gratefully acknowledge support from the Deutsche Forschungsgemeinschaft (DFG) through the project B1 of the SFB 910, and from the European Union’s Horizon	2020 research and innovation program under theSONAR grant agreement no. [734690].


\begin{thebibliography}{63}%
\makeatletter
\providecommand \@ifxundefined [1]{%
 \@ifx{#1\undefined}
}%
\providecommand \@ifnum [1]{%
 \ifnum #1\expandafter \@firstoftwo
 \else \expandafter \@secondoftwo
 \fi
}%
\providecommand \@ifx [1]{%
 \ifx #1\expandafter \@firstoftwo
 \else \expandafter \@secondoftwo
 \fi
}%
\providecommand \natexlab [1]{#1}%
\providecommand \enquote  [1]{``#1''}%
\providecommand \bibnamefont  [1]{#1}%
\providecommand \bibfnamefont [1]{#1}%
\providecommand \citenamefont [1]{#1}%
\providecommand \href@noop [0]{\@secondoftwo}%
\providecommand \href [0]{\begingroup \@sanitize@url \@href}%
\providecommand \@href[1]{\@@startlink{#1}\@@href}%
\providecommand \@@href[1]{\endgroup#1\@@endlink}%
\providecommand \@sanitize@url [0]{\catcode `\\12\catcode `\$12\catcode
  `\&12\catcode `\#12\catcode `\^12\catcode `\_12\catcode `\%12\relax}%
\providecommand \@@startlink[1]{}%
\providecommand \@@endlink[0]{}%
\providecommand \url  [0]{\begingroup\@sanitize@url \@url }%
\providecommand \@url [1]{\endgroup\@href {#1}{\urlprefix }}%
\providecommand \urlprefix  [0]{URL }%
\providecommand \Eprint [0]{\href }%
\providecommand \doibase [0]{https://doi.org/}%
\providecommand \selectlanguage [0]{\@gobble}%
\providecommand \bibinfo  [0]{\@secondoftwo}%
\providecommand \bibfield  [0]{\@secondoftwo}%
\providecommand \translation [1]{[#1]}%
\providecommand \BibitemOpen [0]{}%
\providecommand \bibitemStop [0]{}%
\providecommand \bibitemNoStop [0]{.\EOS\space}%
\providecommand \EOS [0]{\spacefactor3000\relax}%
\providecommand \BibitemShut  [1]{\csname bibitem#1\endcsname}%
\let\auto@bib@innerbib\@empty
\bibitem [{\citenamefont {Kubo}(1966)}]{kubo1966fluctuation}%
  \BibitemOpen
  \bibfield  {author} {\bibinfo {author} {\bibfnamefont {R.}~\bibnamefont
  {Kubo}},\ }\bibfield  {title} {\bibinfo {title} {The fluctuation-dissipation
  theorem},\ }\href@noop {} {\bibfield  {journal} {\bibinfo  {journal} {Reports
  on progress in physics}\ }\textbf {\bibinfo {volume} {29}},\ \bibinfo {pages}
  {255} (\bibinfo {year} {1966})}\BibitemShut {NoStop}%
\bibitem [{\citenamefont {Zwanzig}(2001)}]{zwanzig2001nonequilibrium}%
  \BibitemOpen
  \bibfield  {author} {\bibinfo {author} {\bibfnamefont {R.}~\bibnamefont
  {Zwanzig}},\ }\href@noop {} {\emph {\bibinfo {title} {Nonequilibrium
  statistical mechanics}}}\ (\bibinfo  {publisher} {Oxford University Press},\
  \bibinfo {year} {2001})\BibitemShut {NoStop}%
\bibitem [{\citenamefont {Gardiner}\ \emph {et~al.}(2004)\citenamefont
  {Gardiner}, \citenamefont {Zoller},\ and\ \citenamefont
  {Zoller}}]{gardiner2004quantum}%
  \BibitemOpen
  \bibfield  {author} {\bibinfo {author} {\bibfnamefont {C.}~\bibnamefont
  {Gardiner}}, \bibinfo {author} {\bibfnamefont {P.}~\bibnamefont {Zoller}},\
  and\ \bibinfo {author} {\bibfnamefont {P.}~\bibnamefont {Zoller}},\
  }\href@noop {} {\emph {\bibinfo {title} {Quantum noise: a handbook of
  Markovian and non-Markovian quantum stochastic methods with applications to
  quantum optics}}},\ Vol.~\bibinfo {volume} {56}\ (\bibinfo  {publisher}
  {Springer Science \& Business Media},\ \bibinfo {year} {2004})\BibitemShut
  {NoStop}%
\bibitem [{\citenamefont {Gardiner}(2009)}]{gardiner2009stochastic}%
  \BibitemOpen
  \bibfield  {author} {\bibinfo {author} {\bibfnamefont {C.}~\bibnamefont
  {Gardiner}},\ }\href@noop {} {\emph {\bibinfo {title} {Stochastic
  methods}}},\ Vol.~\bibinfo {volume} {4}\ (\bibinfo  {publisher} {Springer
  Berlin},\ \bibinfo {year} {2009})\BibitemShut {NoStop}%
\bibitem [{\citenamefont {Risken}(1996)}]{risken}%
  \BibitemOpen
  \bibfield  {author} {\bibinfo {author} {\bibfnamefont {H.}~\bibnamefont
  {Risken}},\ }\bibfield  {title} {\bibinfo {title} {Fokker-planck equation},\
  }in\ \href@noop {} {\emph {\bibinfo {booktitle} {The Fokker-Planck
  Equation}}}\ (\bibinfo  {publisher} {Springer},\ \bibinfo {year} {1996})\
  pp.\ \bibinfo {pages} {63--95}\BibitemShut {NoStop}%
\bibitem [{\citenamefont {Nelson}(1967)}]{nelson1967dynamical}%
  \BibitemOpen
  \bibfield  {author} {\bibinfo {author} {\bibfnamefont {E.}~\bibnamefont
  {Nelson}},\ }\href@noop {} {\emph {\bibinfo {title} {Dynamical theories of
  Brownian motion}}},\ Vol.~\bibinfo {volume} {3}\ (\bibinfo  {publisher}
  {Princeton university press},\ \bibinfo {year} {1967})\BibitemShut {NoStop}%
\bibitem [{\citenamefont {Kubo}\ \emph {et~al.}(2012)\citenamefont {Kubo},
  \citenamefont {Toda},\ and\ \citenamefont
  {Hashitsume}}]{kubo2012statistical}%
  \BibitemOpen
  \bibfield  {author} {\bibinfo {author} {\bibfnamefont {R.}~\bibnamefont
  {Kubo}}, \bibinfo {author} {\bibfnamefont {M.}~\bibnamefont {Toda}},\ and\
  \bibinfo {author} {\bibfnamefont {N.}~\bibnamefont {Hashitsume}},\
  }\href@noop {} {\emph {\bibinfo {title} {Statistical physics II:
  nonequilibrium statistical mechanics}}},\ Vol.~\bibinfo {volume} {31}\
  (\bibinfo  {publisher} {Springer Science \& Business Media},\ \bibinfo {year}
  {2012})\BibitemShut {NoStop}%
\bibitem [{\citenamefont {Sekimoto}(2010)}]{sekimoto}%
  \BibitemOpen
  \bibfield  {author} {\bibinfo {author} {\bibfnamefont {K.}~\bibnamefont
  {Sekimoto}},\ }\href@noop {} {\emph {\bibinfo {title} {Stochastic
  energetics}}},\ Vol.\ \bibinfo {volume} {799}\ (\bibinfo  {publisher}
  {Springer},\ \bibinfo {year} {2010})\BibitemShut {NoStop}%
\bibitem [{\citenamefont {Uhlenbeck}\ and\ \citenamefont
  {Ornstein}(1930)}]{uhlenbeck1930theory}%
  \BibitemOpen
  \bibfield  {author} {\bibinfo {author} {\bibfnamefont {G.~E.}\ \bibnamefont
  {Uhlenbeck}}\ and\ \bibinfo {author} {\bibfnamefont {L.~S.}\ \bibnamefont
  {Ornstein}},\ }\bibfield  {title} {\bibinfo {title} {On the theory of the
  brownian motion},\ }\href@noop {} {\bibfield  {journal} {\bibinfo  {journal}
  {Physical review}\ }\textbf {\bibinfo {volume} {36}},\ \bibinfo {pages} {823}
  (\bibinfo {year} {1930})}\BibitemShut {NoStop}%
\bibitem [{\citenamefont {Wax}(1954)}]{wax1954selected}%
  \BibitemOpen
  \bibfield  {author} {\bibinfo {author} {\bibfnamefont {N.}~\bibnamefont
  {Wax}},\ }\href@noop {} {\emph {\bibinfo {title} {Selected papers on noise
  and stochastic processes}}}\ (\bibinfo  {publisher} {Courier Dover
  Publications},\ \bibinfo {year} {1954})\BibitemShut {NoStop}%
\bibitem [{\citenamefont {Wang}\ and\ \citenamefont
  {Uhlenbeck}(1945)}]{wang1945theory}%
  \BibitemOpen
  \bibfield  {author} {\bibinfo {author} {\bibfnamefont {M.~C.}\ \bibnamefont
  {Wang}}\ and\ \bibinfo {author} {\bibfnamefont {G.~E.}\ \bibnamefont
  {Uhlenbeck}},\ }\bibfield  {title} {\bibinfo {title} {On the theory of the
  brownian motion ii},\ }\href@noop {} {\bibfield  {journal} {\bibinfo
  {journal} {Reviews of modern physics}\ }\textbf {\bibinfo {volume} {17}},\
  \bibinfo {pages} {323} (\bibinfo {year} {1945})}\BibitemShut {NoStop}%
\bibitem [{\citenamefont {Einstein}\ \emph {et~al.}(1905)\citenamefont
  {Einstein} \emph {et~al.}}]{einstein1905motion}%
  \BibitemOpen
  \bibfield  {author} {\bibinfo {author} {\bibfnamefont {A.}~\bibnamefont
  {Einstein}} \emph {et~al.},\ }\bibfield  {title} {\bibinfo {title} {On the
  motion of small particles suspended in liquids at rest required by the
  molecular-kinetic theory of heat},\ }\href@noop {} {\bibfield  {journal}
  {\bibinfo  {journal} {Annalen der physik}\ }\textbf {\bibinfo {volume}
  {17}},\ \bibinfo {pages} {549} (\bibinfo {year} {1905})}\BibitemShut
  {NoStop}%
\bibitem [{\citenamefont {Smoluchowski}(1918)}]{smoluchowski1918versuch}%
  \BibitemOpen
  \bibfield  {author} {\bibinfo {author} {\bibfnamefont {M.~v.}\ \bibnamefont
  {Smoluchowski}},\ }\bibfield  {title} {\bibinfo {title} {Versuch einer
  mathematischen theorie der koagulationskinetik kolloider l{\"o}sungen},\
  }\href@noop {} {\bibfield  {journal} {\bibinfo  {journal} {Zeitschrift
  f{\"u}r physikalische Chemie}\ }\textbf {\bibinfo {volume} {92}},\ \bibinfo
  {pages} {129} (\bibinfo {year} {1918})}\BibitemShut {NoStop}%
\bibitem [{\citenamefont {Philipse}(2018)}]{philipse2018}%
  \BibitemOpen
  \bibfield  {author} {\bibinfo {author} {\bibfnamefont {A.~P.}\ \bibnamefont
  {Philipse}},\ }\href@noop {} {\emph {\bibinfo {title} {Brownian Motion}}}\
  (\bibinfo  {publisher} {Springer},\ \bibinfo {year} {2018})\BibitemShut
  {NoStop}%
\bibitem [{\citenamefont {Huang}\ \emph {et~al.}(2011)\citenamefont {Huang},
  \citenamefont {Chavez}, \citenamefont {Taute}, \citenamefont {Luki{\'c}},
  \citenamefont {Jeney}, \citenamefont {Raizen},\ and\ \citenamefont
  {Florin}}]{huang2011direct}%
  \BibitemOpen
  \bibfield  {author} {\bibinfo {author} {\bibfnamefont {R.}~\bibnamefont
  {Huang}}, \bibinfo {author} {\bibfnamefont {I.}~\bibnamefont {Chavez}},
  \bibinfo {author} {\bibfnamefont {K.~M.}\ \bibnamefont {Taute}}, \bibinfo
  {author} {\bibfnamefont {B.}~\bibnamefont {Luki{\'c}}}, \bibinfo {author}
  {\bibfnamefont {S.}~\bibnamefont {Jeney}}, \bibinfo {author} {\bibfnamefont
  {M.~G.}\ \bibnamefont {Raizen}},\ and\ \bibinfo {author} {\bibfnamefont
  {E.-L.}\ \bibnamefont {Florin}},\ }\bibfield  {title} {\bibinfo {title}
  {Direct observation of the full transition from ballistic to diffusive
  brownian motion in a liquid},\ }\href@noop {} {\bibfield  {journal} {\bibinfo
   {journal} {Nature Physics}\ }\textbf {\bibinfo {volume} {7}},\ \bibinfo
  {pages} {576} (\bibinfo {year} {2011})}\BibitemShut {NoStop}%
\bibitem [{\citenamefont {Shoghl}\ \emph {et~al.}(2016)\citenamefont {Shoghl},
  \citenamefont {Jamali},\ and\ \citenamefont
  {Moraveji}}]{shoghl2016electrical}%
  \BibitemOpen
  \bibfield  {author} {\bibinfo {author} {\bibfnamefont {S.~N.}\ \bibnamefont
  {Shoghl}}, \bibinfo {author} {\bibfnamefont {J.}~\bibnamefont {Jamali}},\
  and\ \bibinfo {author} {\bibfnamefont {M.~K.}\ \bibnamefont {Moraveji}},\
  }\bibfield  {title} {\bibinfo {title} {Electrical conductivity, viscosity,
  and density of different nanofluids: An experimental study},\ }\href@noop {}
  {\bibfield  {journal} {\bibinfo  {journal} {Experimental Thermal and Fluid
  Science}\ }\textbf {\bibinfo {volume} {74}},\ \bibinfo {pages} {339}
  (\bibinfo {year} {2016})}\BibitemShut {NoStop}%
\bibitem [{\citenamefont {Saffman}\ and\ \citenamefont
  {Delbr{\"u}ck}(1975)}]{saffman1975brownian}%
  \BibitemOpen
  \bibfield  {author} {\bibinfo {author} {\bibfnamefont {P.}~\bibnamefont
  {Saffman}}\ and\ \bibinfo {author} {\bibfnamefont {M.}~\bibnamefont
  {Delbr{\"u}ck}},\ }\bibfield  {title} {\bibinfo {title} {Brownian motion in
  biological membranes},\ }\href@noop {} {\bibfield  {journal} {\bibinfo
  {journal} {Proceedings of the National Academy of Sciences}\ }\textbf
  {\bibinfo {volume} {72}},\ \bibinfo {pages} {3111} (\bibinfo {year}
  {1975})}\BibitemShut {NoStop}%
\bibitem [{\citenamefont {Pusey}(1978)}]{pusey1978intensity}%
  \BibitemOpen
  \bibfield  {author} {\bibinfo {author} {\bibfnamefont {P.~N.}\ \bibnamefont
  {Pusey}},\ }\bibfield  {title} {\bibinfo {title} {Intensity fluctuation
  spectroscopy of charged brownian particles: the coherent scattering
  function},\ }\href@noop {} {\bibfield  {journal} {\bibinfo  {journal}
  {Journal of Physics A: Mathematical and General}\ }\textbf {\bibinfo {volume}
  {11}},\ \bibinfo {pages} {119} (\bibinfo {year} {1978})}\BibitemShut
  {NoStop}%
\bibitem [{\citenamefont {Maret}\ and\ \citenamefont
  {Wolf}(1987)}]{maret1987multiple}%
  \BibitemOpen
  \bibfield  {author} {\bibinfo {author} {\bibfnamefont {G.}~\bibnamefont
  {Maret}}\ and\ \bibinfo {author} {\bibfnamefont {P.}~\bibnamefont {Wolf}},\
  }\bibfield  {title} {\bibinfo {title} {Multiple light scattering from
  disordered media. the effect of brownian motion of scatterers},\ }\href@noop
  {} {\bibfield  {journal} {\bibinfo  {journal} {Zeitschrift f{\"u}r Physik B
  Condensed Matter}\ }\textbf {\bibinfo {volume} {65}},\ \bibinfo {pages} {409}
  (\bibinfo {year} {1987})}\BibitemShut {NoStop}%
\bibitem [{\citenamefont {Burada}\ \emph {et~al.}(2009)\citenamefont {Burada},
  \citenamefont {H{\"a}nggi}, \citenamefont {Marchesoni}, \citenamefont
  {Schmid},\ and\ \citenamefont {Talkner}}]{burada2009diffusion}%
  \BibitemOpen
  \bibfield  {author} {\bibinfo {author} {\bibfnamefont {P.~S.}\ \bibnamefont
  {Burada}}, \bibinfo {author} {\bibfnamefont {P.}~\bibnamefont {H{\"a}nggi}},
  \bibinfo {author} {\bibfnamefont {F.}~\bibnamefont {Marchesoni}}, \bibinfo
  {author} {\bibfnamefont {G.}~\bibnamefont {Schmid}},\ and\ \bibinfo {author}
  {\bibfnamefont {P.}~\bibnamefont {Talkner}},\ }\bibfield  {title} {\bibinfo
  {title} {Diffusion in confined geometries},\ }\href@noop {} {\bibfield
  {journal} {\bibinfo  {journal} {ChemPhysChem}\ }\textbf {\bibinfo {volume}
  {10}},\ \bibinfo {pages} {45} (\bibinfo {year} {2009})}\BibitemShut {NoStop}%
\bibitem [{\citenamefont {Ishii}\ \emph {et~al.}(2005)\citenamefont {Ishii},
  \citenamefont {Yoshida},\ and\ \citenamefont {Iwai}}]{ishii2005single}%
  \BibitemOpen
  \bibfield  {author} {\bibinfo {author} {\bibfnamefont {K.}~\bibnamefont
  {Ishii}}, \bibinfo {author} {\bibfnamefont {R.}~\bibnamefont {Yoshida}},\
  and\ \bibinfo {author} {\bibfnamefont {T.}~\bibnamefont {Iwai}},\ }\bibfield
  {title} {\bibinfo {title} {Single-scattering spectroscopy for extremely dense
  colloidal suspensions by use of a low-coherence interferometer},\ }\href@noop
  {} {\bibfield  {journal} {\bibinfo  {journal} {Optics letters}\ }\textbf
  {\bibinfo {volume} {30}},\ \bibinfo {pages} {555} (\bibinfo {year}
  {2005})}\BibitemShut {NoStop}%
\bibitem [{\citenamefont {Gardiner}\ and\ \citenamefont
  {Zoller}(2015)}]{gardiner2015quantum}%
  \BibitemOpen
  \bibfield  {author} {\bibinfo {author} {\bibfnamefont {C.}~\bibnamefont
  {Gardiner}}\ and\ \bibinfo {author} {\bibfnamefont {P.}~\bibnamefont
  {Zoller}},\ }\href@noop {} {\emph {\bibinfo {title} {The quantum world of
  ultra-cold atoms and light book II: the physics of quantum-optical
  devices}}},\ Vol.~\bibinfo {volume} {4}\ (\bibinfo  {publisher} {World
  Scientific Publishing Company},\ \bibinfo {year} {2015})\BibitemShut
  {NoStop}%
\bibitem [{\citenamefont {Hoi}\ \emph {et~al.}(2015)\citenamefont {Hoi},
  \citenamefont {Kockum}, \citenamefont {Tornberg}, \citenamefont
  {Pourkabirian}, \citenamefont {Johansson}, \citenamefont {Delsing},\ and\
  \citenamefont {Wilson}}]{hoi2015probing}%
  \BibitemOpen
  \bibfield  {author} {\bibinfo {author} {\bibfnamefont {I.-C.}\ \bibnamefont
  {Hoi}}, \bibinfo {author} {\bibfnamefont {A.}~\bibnamefont {Kockum}},
  \bibinfo {author} {\bibfnamefont {L.}~\bibnamefont {Tornberg}}, \bibinfo
  {author} {\bibfnamefont {A.}~\bibnamefont {Pourkabirian}}, \bibinfo {author}
  {\bibfnamefont {G.}~\bibnamefont {Johansson}}, \bibinfo {author}
  {\bibfnamefont {P.}~\bibnamefont {Delsing}},\ and\ \bibinfo {author}
  {\bibfnamefont {C.}~\bibnamefont {Wilson}},\ }\bibfield  {title} {\bibinfo
  {title} {Probing the quantum vacuum with an artificial atom in front of a
  mirror},\ }\href@noop {} {\bibfield  {journal} {\bibinfo  {journal} {Nature
  Physics}\ }\textbf {\bibinfo {volume} {11}},\ \bibinfo {pages} {1045}
  (\bibinfo {year} {2015})}\BibitemShut {NoStop}%
\bibitem [{\citenamefont {Svidzinsky}\ \emph {et~al.}(2018)\citenamefont
  {Svidzinsky}, \citenamefont {Ben-Benjamin}, \citenamefont {Fulling},\ and\
  \citenamefont {Page}}]{svidzinsky2018excitation}%
  \BibitemOpen
  \bibfield  {author} {\bibinfo {author} {\bibfnamefont {A.~A.}\ \bibnamefont
  {Svidzinsky}}, \bibinfo {author} {\bibfnamefont {J.~S.}\ \bibnamefont
  {Ben-Benjamin}}, \bibinfo {author} {\bibfnamefont {S.~A.}\ \bibnamefont
  {Fulling}},\ and\ \bibinfo {author} {\bibfnamefont {D.~N.}\ \bibnamefont
  {Page}},\ }\bibfield  {title} {\bibinfo {title} {Excitation of an atom by a
  uniformly accelerated mirror through virtual transitions},\ }\href@noop {}
  {\bibfield  {journal} {\bibinfo  {journal} {Physical review letters}\
  }\textbf {\bibinfo {volume} {121}},\ \bibinfo {pages} {071301} (\bibinfo
  {year} {2018})}\BibitemShut {NoStop}%
\bibitem [{\citenamefont {H{\'e}tet}\ \emph {et~al.}(2011)\citenamefont
  {H{\'e}tet}, \citenamefont {Slodi{\v{c}}ka}, \citenamefont {Hennrich},\ and\
  \citenamefont {Blatt}}]{hetet2011single}%
  \BibitemOpen
  \bibfield  {author} {\bibinfo {author} {\bibfnamefont {G.}~\bibnamefont
  {H{\'e}tet}}, \bibinfo {author} {\bibfnamefont {L.}~\bibnamefont
  {Slodi{\v{c}}ka}}, \bibinfo {author} {\bibfnamefont {M.}~\bibnamefont
  {Hennrich}},\ and\ \bibinfo {author} {\bibfnamefont {R.}~\bibnamefont
  {Blatt}},\ }\bibfield  {title} {\bibinfo {title} {Single atom as a mirror of
  an optical cavity},\ }\href@noop {} {\bibfield  {journal} {\bibinfo
  {journal} {Physical review letters}\ }\textbf {\bibinfo {volume} {107}},\
  \bibinfo {pages} {133002} (\bibinfo {year} {2011})}\BibitemShut {NoStop}%
\bibitem [{\citenamefont {Wilson}\ \emph {et~al.}(2003)\citenamefont {Wilson},
  \citenamefont {Bushev}, \citenamefont {Eschner}, \citenamefont
  {Schmidt-Kaler}, \citenamefont {Becher}, \citenamefont {Blatt},\ and\
  \citenamefont {Dorner}}]{wilson2003vacuum}%
  \BibitemOpen
  \bibfield  {author} {\bibinfo {author} {\bibfnamefont {M.}~\bibnamefont
  {Wilson}}, \bibinfo {author} {\bibfnamefont {P.}~\bibnamefont {Bushev}},
  \bibinfo {author} {\bibfnamefont {J.}~\bibnamefont {Eschner}}, \bibinfo
  {author} {\bibfnamefont {F.}~\bibnamefont {Schmidt-Kaler}}, \bibinfo {author}
  {\bibfnamefont {C.}~\bibnamefont {Becher}}, \bibinfo {author} {\bibfnamefont
  {R.}~\bibnamefont {Blatt}},\ and\ \bibinfo {author} {\bibfnamefont
  {U.}~\bibnamefont {Dorner}},\ }\bibfield  {title} {\bibinfo {title}
  {Vacuum-field level shifts in a single trapped ion mediated by a single
  distant mirror},\ }\href@noop {} {\bibfield  {journal} {\bibinfo  {journal}
  {Physical review letters}\ }\textbf {\bibinfo {volume} {91}},\ \bibinfo
  {pages} {213602} (\bibinfo {year} {2003})}\BibitemShut {NoStop}%
\bibitem [{\citenamefont {Glaetzle}\ \emph {et~al.}(2010)\citenamefont
  {Glaetzle}, \citenamefont {Hammerer}, \citenamefont {Daley}, \citenamefont
  {Blatt},\ and\ \citenamefont {Zoller}}]{glaetzle2010single}%
  \BibitemOpen
  \bibfield  {author} {\bibinfo {author} {\bibfnamefont {A.~W.}\ \bibnamefont
  {Glaetzle}}, \bibinfo {author} {\bibfnamefont {K.}~\bibnamefont {Hammerer}},
  \bibinfo {author} {\bibfnamefont {A.}~\bibnamefont {Daley}}, \bibinfo
  {author} {\bibfnamefont {R.}~\bibnamefont {Blatt}},\ and\ \bibinfo {author}
  {\bibfnamefont {P.}~\bibnamefont {Zoller}},\ }\bibfield  {title} {\bibinfo
  {title} {A single trapped atom in front of an oscillating mirror},\
  }\href@noop {} {\bibfield  {journal} {\bibinfo  {journal} {Optics
  Communications}\ }\textbf {\bibinfo {volume} {283}},\ \bibinfo {pages} {758}
  (\bibinfo {year} {2010})}\BibitemShut {NoStop}%
\bibitem [{\citenamefont {Dubin}\ \emph {et~al.}(2007)\citenamefont {Dubin},
  \citenamefont {Rotter}, \citenamefont {Mukherjee}, \citenamefont {Russo},
  \citenamefont {Eschner},\ and\ \citenamefont {Blatt}}]{dubin2007photon}%
  \BibitemOpen
  \bibfield  {author} {\bibinfo {author} {\bibfnamefont {F.}~\bibnamefont
  {Dubin}}, \bibinfo {author} {\bibfnamefont {D.}~\bibnamefont {Rotter}},
  \bibinfo {author} {\bibfnamefont {M.}~\bibnamefont {Mukherjee}}, \bibinfo
  {author} {\bibfnamefont {C.}~\bibnamefont {Russo}}, \bibinfo {author}
  {\bibfnamefont {J.}~\bibnamefont {Eschner}},\ and\ \bibinfo {author}
  {\bibfnamefont {R.}~\bibnamefont {Blatt}},\ }\bibfield  {title} {\bibinfo
  {title} {Photon correlation versus interference of single-atom fluorescence
  in a half-cavity},\ }\href@noop {} {\bibfield  {journal} {\bibinfo  {journal}
  {Physical review letters}\ }\textbf {\bibinfo {volume} {98}},\ \bibinfo
  {pages} {183003} (\bibinfo {year} {2007})}\BibitemShut {NoStop}%
\bibitem [{\citenamefont {Weisskopf}\ and\ \citenamefont
  {Wigner}(1997)}]{weisskopf1997naturliche}%
  \BibitemOpen
  \bibfield  {author} {\bibinfo {author} {\bibfnamefont {V.}~\bibnamefont
  {Weisskopf}}\ and\ \bibinfo {author} {\bibfnamefont {E.}~\bibnamefont
  {Wigner}},\ }\bibfield  {title} {\bibinfo {title} {{\"U}ber die
  nat{\"u}rliche linienbreite in der strahlung des harmonischen oszillators},\
  }in\ \href@noop {} {\emph {\bibinfo {booktitle} {Part I: Particles and
  Fields. Part II: Foundations of Quantum Mechanics}}}\ (\bibinfo  {publisher}
  {Springer},\ \bibinfo {year} {1997})\ pp.\ \bibinfo {pages}
  {50--61}\BibitemShut {NoStop}%
\bibitem [{\citenamefont {Dorner}\ and\ \citenamefont
  {Zoller}(2002)}]{dorner-half-cavities}%
  \BibitemOpen
  \bibfield  {author} {\bibinfo {author} {\bibfnamefont {U.}~\bibnamefont
  {Dorner}}\ and\ \bibinfo {author} {\bibfnamefont {P.}~\bibnamefont
  {Zoller}},\ }\bibfield  {title} {\bibinfo {title} {Laser-driven atoms in
  half-cavities},\ }\href@noop {} {\bibfield  {journal} {\bibinfo  {journal}
  {Phys. Rev. A}\ }\textbf {\bibinfo {volume} {66}},\ \bibinfo {pages} {023816}
  (\bibinfo {year} {2002})}\BibitemShut {NoStop}%
\bibitem [{\citenamefont {Cook}\ and\ \citenamefont
  {Milonni}(1987{\natexlab{a}})}]{Cook86feedback}%
  \BibitemOpen
  \bibfield  {author} {\bibinfo {author} {\bibfnamefont {R.~J.}\ \bibnamefont
  {Cook}}\ and\ \bibinfo {author} {\bibfnamefont {P.~W.}\ \bibnamefont
  {Milonni}},\ }\bibfield  {title} {\bibinfo {title} {{Quantum theory of an
  atom near partially reflecting walls}},\ }\href@noop {} {\bibfield  {journal}
  {\bibinfo  {journal} {Phys. Rev. A}\ }\textbf {\bibinfo {volume} {35}},\
  \bibinfo {pages} {5081} (\bibinfo {year} {1987}{\natexlab{a}})}\BibitemShut
  {NoStop}%
\bibitem [{\citenamefont {Milonni}\ and\ \citenamefont
  {Knight}(1974)}]{milonni}%
  \BibitemOpen
  \bibfield  {author} {\bibinfo {author} {\bibfnamefont {P.~W.}\ \bibnamefont
  {Milonni}}\ and\ \bibinfo {author} {\bibfnamefont {P.~L.}\ \bibnamefont
  {Knight}},\ }\bibfield  {title} {\bibinfo {title} {Retardation in the
  resonant interaction of two identical atoms},\ }\href
  {https://doi.org/10.1103/PhysRevA.10.1096} {\bibfield  {journal} {\bibinfo
  {journal} {Phys. Rev. A}\ }\textbf {\bibinfo {volume} {10}},\ \bibinfo
  {pages} {1096} (\bibinfo {year} {1974})}\BibitemShut {NoStop}%
\bibitem [{\citenamefont {Cook}\ and\ \citenamefont
  {Milonni}(1987{\natexlab{b}})}]{cook1987quantum}%
  \BibitemOpen
  \bibfield  {author} {\bibinfo {author} {\bibfnamefont {R.}~\bibnamefont
  {Cook}}\ and\ \bibinfo {author} {\bibfnamefont {P.}~\bibnamefont {Milonni}},\
  }\bibfield  {title} {\bibinfo {title} {Quantum theory of an atom near
  partially reflecting walls},\ }\href@noop {} {\bibfield  {journal} {\bibinfo
  {journal} {Physical Review A}\ }\textbf {\bibinfo {volume} {35}},\ \bibinfo
  {pages} {5081} (\bibinfo {year} {1987}{\natexlab{b}})}\BibitemShut {NoStop}%
\bibitem [{\citenamefont {Cook}\ \emph {et~al.}(2018)\citenamefont {Cook},
  \citenamefont {Schuster}, \citenamefont {Cleland},\ and\ \citenamefont
  {Jacobs}}]{cook2018input}%
  \BibitemOpen
  \bibfield  {author} {\bibinfo {author} {\bibfnamefont {R.}~\bibnamefont
  {Cook}}, \bibinfo {author} {\bibfnamefont {D.}~\bibnamefont {Schuster}},
  \bibinfo {author} {\bibfnamefont {A.}~\bibnamefont {Cleland}},\ and\ \bibinfo
  {author} {\bibfnamefont {K.}~\bibnamefont {Jacobs}},\ }\bibfield  {title}
  {\bibinfo {title} {Input-output theory for superconducting and photonic
  circuits that contain weak retro-reflections and other weak
  pseudo-cavities},\ }\href@noop {} {\bibfield  {journal} {\bibinfo  {journal}
  {arXiv preprint arXiv:1803.04763}\ } (\bibinfo {year} {2018})}\BibitemShut
  {NoStop}%
\bibitem [{\citenamefont {N{\'e}met}\ and\ \citenamefont
  {Parkins}(2016)}]{nemet2016enhanced}%
  \BibitemOpen
  \bibfield  {author} {\bibinfo {author} {\bibfnamefont {N.}~\bibnamefont
  {N{\'e}met}}\ and\ \bibinfo {author} {\bibfnamefont {S.}~\bibnamefont
  {Parkins}},\ }\bibfield  {title} {\bibinfo {title} {Enhanced optical
  squeezing from a degenerate parametric amplifier via time-delayed coherent
  feedback},\ }\href@noop {} {\bibfield  {journal} {\bibinfo  {journal}
  {Physical Review A}\ }\textbf {\bibinfo {volume} {94}},\ \bibinfo {pages}
  {023809} (\bibinfo {year} {2016})}\BibitemShut {NoStop}%
\bibitem [{\citenamefont {Faulstich}\ \emph {et~al.}(2018)\citenamefont
  {Faulstich}, \citenamefont {Kraft},\ and\ \citenamefont
  {Carmele}}]{faulstich2018unraveling}%
  \BibitemOpen
  \bibfield  {author} {\bibinfo {author} {\bibfnamefont {F.~M.}\ \bibnamefont
  {Faulstich}}, \bibinfo {author} {\bibfnamefont {M.}~\bibnamefont {Kraft}},\
  and\ \bibinfo {author} {\bibfnamefont {A.}~\bibnamefont {Carmele}},\
  }\bibfield  {title} {\bibinfo {title} {Unraveling mirror properties in
  time-delayed quantum feedback scenarios},\ }\href@noop {} {\bibfield
  {journal} {\bibinfo  {journal} {Journal of Modern Optics}\ }\textbf {\bibinfo
  {volume} {65}},\ \bibinfo {pages} {1323} (\bibinfo {year}
  {2018})}\BibitemShut {NoStop}%
\bibitem [{\citenamefont {Pichler}\ and\ \citenamefont
  {Zoller}(2015)}]{pichler2015photonic}%
  \BibitemOpen
  \bibfield  {author} {\bibinfo {author} {\bibfnamefont {H.}~\bibnamefont
  {Pichler}}\ and\ \bibinfo {author} {\bibfnamefont {P.}~\bibnamefont
  {Zoller}},\ }\bibfield  {title} {\bibinfo {title} {Photonic quantum circuits
  with time delays},\ }\href@noop {} {\bibfield  {journal} {\bibinfo  {journal}
  {arXiv preprint arXiv:1510.04646}\ } (\bibinfo {year} {2015})}\BibitemShut
  {NoStop}%
\bibitem [{\citenamefont {Crowder}\ \emph {et~al.}(2019)\citenamefont
  {Crowder}, \citenamefont {Carmichael},\ and\ \citenamefont
  {Hughes}}]{crowder2019quantum}%
  \BibitemOpen
  \bibfield  {author} {\bibinfo {author} {\bibfnamefont {G.}~\bibnamefont
  {Crowder}}, \bibinfo {author} {\bibfnamefont {H.}~\bibnamefont
  {Carmichael}},\ and\ \bibinfo {author} {\bibfnamefont {S.}~\bibnamefont
  {Hughes}},\ }\bibfield  {title} {\bibinfo {title} {Quantum trajectory theory
  of few photon cavity-qed systems with a time-delayed coherent feedback},\
  }\href@noop {} {\bibfield  {journal} {\bibinfo  {journal} {arXiv preprint
  arXiv:1910.14601}\ } (\bibinfo {year} {2019})}\BibitemShut {NoStop}%
\bibitem [{\citenamefont {Barkemeyer}\ \emph {et~al.}(2019)\citenamefont
  {Barkemeyer}, \citenamefont {Finsterh{\"o}lzl}, \citenamefont {Knorr},\ and\
  \citenamefont {Carmele}}]{barkemeyer2019revisiting}%
  \BibitemOpen
  \bibfield  {author} {\bibinfo {author} {\bibfnamefont {K.}~\bibnamefont
  {Barkemeyer}}, \bibinfo {author} {\bibfnamefont {R.}~\bibnamefont
  {Finsterh{\"o}lzl}}, \bibinfo {author} {\bibfnamefont {A.}~\bibnamefont
  {Knorr}},\ and\ \bibinfo {author} {\bibfnamefont {A.}~\bibnamefont
  {Carmele}},\ }\bibfield  {title} {\bibinfo {title} {Revisiting quantum
  feedback control: Disentangling the feedback-induced phase from the
  corresponding amplitude},\ }\href@noop {} {\bibfield  {journal} {\bibinfo
  {journal} {Advanced Quantum Technologies}\ } (\bibinfo {year}
  {2019})}\BibitemShut {NoStop}%
\bibitem [{\citenamefont {Sch{\"o}ll}\ and\ \citenamefont
  {Schuster}(2008)}]{ChaosControl2008}%
  \BibitemOpen
  \bibfield  {author} {\bibinfo {author} {\bibfnamefont {E.}~\bibnamefont
  {Sch{\"o}ll}}\ and\ \bibinfo {author} {\bibfnamefont {H.~G.}\ \bibnamefont
  {Schuster}},\ }\href@noop {} {\emph {\bibinfo {title} {{Handbook of Chaos
  Control}}}}\ (\bibinfo  {publisher} {Wiley-VCH},\ \bibinfo {year}
  {2008})\BibitemShut {NoStop}%
\bibitem [{\citenamefont {Sch{\"o}ll}\ \emph {et~al.}(2016)\citenamefont
  {Sch{\"o}ll}, \citenamefont {Klapp},\ and\ \citenamefont
  {H{\"o}vel}}]{sfb910}%
  \BibitemOpen
  \bibinfo {editor} {\bibfnamefont {E.}~\bibnamefont {Sch{\"o}ll}}, \bibinfo
  {editor} {\bibfnamefont {S.~H.~L.}\ \bibnamefont {Klapp}},\ and\ \bibinfo
  {editor} {\bibfnamefont {P.}~\bibnamefont {H{\"o}vel}},\ eds.,\ \href@noop {}
  {\emph {\bibinfo {title} {Control of Self-Organizing Nonlinear Systems}}}\
  (\bibinfo  {publisher} {Springer International Publishing},\ \bibinfo {year}
  {2016})\BibitemShut {NoStop}%
\bibitem [{\citenamefont {Droenner}\ \emph {et~al.}(2019)\citenamefont
  {Droenner}, \citenamefont {Naumann}, \citenamefont {Sch{\"o}ll},
  \citenamefont {Knorr},\ and\ \citenamefont {Carmele}}]{droenner2019quantum}%
  \BibitemOpen
  \bibfield  {author} {\bibinfo {author} {\bibfnamefont {L.}~\bibnamefont
  {Droenner}}, \bibinfo {author} {\bibfnamefont {N.~L.}\ \bibnamefont
  {Naumann}}, \bibinfo {author} {\bibfnamefont {E.}~\bibnamefont {Sch{\"o}ll}},
  \bibinfo {author} {\bibfnamefont {A.}~\bibnamefont {Knorr}},\ and\ \bibinfo
  {author} {\bibfnamefont {A.}~\bibnamefont {Carmele}},\ }\bibfield  {title}
  {\bibinfo {title} {Quantum pyragas control: selective control of individual
  photon probabilities},\ }\href@noop {} {\bibfield  {journal} {\bibinfo
  {journal} {Physical Review A}\ }\textbf {\bibinfo {volume} {99}},\ \bibinfo
  {pages} {023840} (\bibinfo {year} {2019})}\BibitemShut {NoStop}%
\bibitem [{\citenamefont {Lu}\ \emph {et~al.}(2017)\citenamefont {Lu},
  \citenamefont {Naumann}, \citenamefont {Cerrillo}, \citenamefont {Zhao},
  \citenamefont {Knorr},\ and\ \citenamefont {Carmele}}]{lu2017intensified}%
  \BibitemOpen
  \bibfield  {author} {\bibinfo {author} {\bibfnamefont {Y.}~\bibnamefont
  {Lu}}, \bibinfo {author} {\bibfnamefont {N.~L.}\ \bibnamefont {Naumann}},
  \bibinfo {author} {\bibfnamefont {J.}~\bibnamefont {Cerrillo}}, \bibinfo
  {author} {\bibfnamefont {Q.}~\bibnamefont {Zhao}}, \bibinfo {author}
  {\bibfnamefont {A.}~\bibnamefont {Knorr}},\ and\ \bibinfo {author}
  {\bibfnamefont {A.}~\bibnamefont {Carmele}},\ }\bibfield  {title} {\bibinfo
  {title} {Intensified antibunching via feedback-induced quantum
  interference},\ }\href@noop {} {\bibfield  {journal} {\bibinfo  {journal}
  {Physical Review A}\ }\textbf {\bibinfo {volume} {95}},\ \bibinfo {pages}
  {063840} (\bibinfo {year} {2017})}\BibitemShut {NoStop}%
\bibitem [{\citenamefont {Guimond}\ \emph {et~al.}(2017)\citenamefont
  {Guimond}, \citenamefont {Pletyukhov}, \citenamefont {Pichler},\ and\
  \citenamefont {Zoller}}]{guimond2017delayed}%
  \BibitemOpen
  \bibfield  {author} {\bibinfo {author} {\bibfnamefont {P.-O.}\ \bibnamefont
  {Guimond}}, \bibinfo {author} {\bibfnamefont {M.}~\bibnamefont {Pletyukhov}},
  \bibinfo {author} {\bibfnamefont {H.}~\bibnamefont {Pichler}},\ and\ \bibinfo
  {author} {\bibfnamefont {P.}~\bibnamefont {Zoller}},\ }\bibfield  {title}
  {\bibinfo {title} {Delayed coherent quantum feedback from a scattering theory
  and a matrix product state perspective},\ }\href@noop {} {\bibfield
  {journal} {\bibinfo  {journal} {Quantum Science and Technology}\ }\textbf
  {\bibinfo {volume} {2}},\ \bibinfo {pages} {044012} (\bibinfo {year}
  {2017})}\BibitemShut {NoStop}%
\bibitem [{\citenamefont {Guimond}\ \emph {et~al.}(2016)\citenamefont
  {Guimond}, \citenamefont {Pichler}, \citenamefont {Rauschenbeutel},\ and\
  \citenamefont {Zoller}}]{guimond2016chiral}%
  \BibitemOpen
  \bibfield  {author} {\bibinfo {author} {\bibfnamefont {P.-O.}\ \bibnamefont
  {Guimond}}, \bibinfo {author} {\bibfnamefont {H.}~\bibnamefont {Pichler}},
  \bibinfo {author} {\bibfnamefont {A.}~\bibnamefont {Rauschenbeutel}},\ and\
  \bibinfo {author} {\bibfnamefont {P.}~\bibnamefont {Zoller}},\ }\bibfield
  {title} {\bibinfo {title} {Chiral quantum optics with v-level atoms and
  coherent quantum feedback},\ }\href@noop {} {\bibfield  {journal} {\bibinfo
  {journal} {Physical Review A}\ }\textbf {\bibinfo {volume} {94}},\ \bibinfo
  {pages} {033829} (\bibinfo {year} {2016})}\BibitemShut {NoStop}%
\bibitem [{\citenamefont {Kabuss}\ \emph {et~al.}(2015)\citenamefont {Kabuss},
  \citenamefont {Krimer}, \citenamefont {Rotter}, \citenamefont {Stannigel},
  \citenamefont {Knorr},\ and\ \citenamefont {Carmele}}]{kabuss2015analytical}%
  \BibitemOpen
  \bibfield  {author} {\bibinfo {author} {\bibfnamefont {J.}~\bibnamefont
  {Kabuss}}, \bibinfo {author} {\bibfnamefont {D.~O.}\ \bibnamefont {Krimer}},
  \bibinfo {author} {\bibfnamefont {S.}~\bibnamefont {Rotter}}, \bibinfo
  {author} {\bibfnamefont {K.}~\bibnamefont {Stannigel}}, \bibinfo {author}
  {\bibfnamefont {A.}~\bibnamefont {Knorr}},\ and\ \bibinfo {author}
  {\bibfnamefont {A.}~\bibnamefont {Carmele}},\ }\bibfield  {title} {\bibinfo
  {title} {Analytical study of quantum feedback enhanced rabi oscillations},\
  }\href@noop {} {\bibfield  {journal} {\bibinfo  {journal} {arXiv preprint
  arXiv:1503.05722}\ } (\bibinfo {year} {2015})}\BibitemShut {NoStop}%
\bibitem [{\citenamefont {N\'emet}\ \emph {et~al.}(2019)\citenamefont
  {N\'emet}, \citenamefont {Carmele}, \citenamefont {Parkins},\ and\
  \citenamefont {Knorr}}]{multimode}%
  \BibitemOpen
  \bibfield  {author} {\bibinfo {author} {\bibfnamefont {N.}~\bibnamefont
  {N\'emet}}, \bibinfo {author} {\bibfnamefont {A.}~\bibnamefont {Carmele}},
  \bibinfo {author} {\bibfnamefont {S.}~\bibnamefont {Parkins}},\ and\ \bibinfo
  {author} {\bibfnamefont {A.}~\bibnamefont {Knorr}},\ }\bibfield  {title}
  {\bibinfo {title} {Comparison between continuous- and discrete-mode coherent
  feedback for the jaynes-cummings model},\ }\href
  {https://doi.org/10.1103/PhysRevA.100.023805} {\bibfield  {journal} {\bibinfo
   {journal} {Phys. Rev. A}\ }\textbf {\bibinfo {volume} {100}},\ \bibinfo
  {pages} {023805} (\bibinfo {year} {2019})}\BibitemShut {NoStop}%
\bibitem [{\citenamefont {Carmele}\ \emph {et~al.}(2019)\citenamefont
  {Carmele}, \citenamefont {Nemet}, \citenamefont {Canela},\ and\ \citenamefont
  {Parkins}}]{carmele2019retardation}%
  \BibitemOpen
  \bibfield  {author} {\bibinfo {author} {\bibfnamefont {A.}~\bibnamefont
  {Carmele}}, \bibinfo {author} {\bibfnamefont {N.}~\bibnamefont {Nemet}},
  \bibinfo {author} {\bibfnamefont {V.}~\bibnamefont {Canela}},\ and\ \bibinfo
  {author} {\bibfnamefont {S.}~\bibnamefont {Parkins}},\ }\bibfield  {title}
  {\bibinfo {title} {Retardation-induced anomalous population trapping in a
  multiply-excited, multiple-emitter waveguide-qed system},\ }\href@noop {}
  {\bibfield  {journal} {\bibinfo  {journal} {arXiv preprint arXiv:1910.13414}\
  } (\bibinfo {year} {2019})}\BibitemShut {NoStop}%
\bibitem [{\citenamefont {Kabuss}\ \emph {et~al.}(2016)\citenamefont {Kabuss},
  \citenamefont {Katsch}, \citenamefont {Knorr},\ and\ \citenamefont
  {Carmele}}]{kabuss2016}%
  \BibitemOpen
  \bibfield  {author} {\bibinfo {author} {\bibfnamefont {J.}~\bibnamefont
  {Kabuss}}, \bibinfo {author} {\bibfnamefont {F.}~\bibnamefont {Katsch}},
  \bibinfo {author} {\bibfnamefont {A.}~\bibnamefont {Knorr}},\ and\ \bibinfo
  {author} {\bibfnamefont {A.}~\bibnamefont {Carmele}},\ }\bibfield  {title}
  {\bibinfo {title} {Unraveling coherent quantum feedback for pyragas
  control},\ }\href@noop {} {\bibfield  {journal} {\bibinfo  {journal} {JOSA
  B}\ }\textbf {\bibinfo {volume} {33}},\ \bibinfo {pages} {C10} (\bibinfo
  {year} {2016})}\BibitemShut {NoStop}%
\bibitem [{\citenamefont {H{\"a}nggi}\ and\ \citenamefont
  {Ingold}(2005)}]{hanggi2005fundamental}%
  \BibitemOpen
  \bibfield  {author} {\bibinfo {author} {\bibfnamefont {P.}~\bibnamefont
  {H{\"a}nggi}}\ and\ \bibinfo {author} {\bibfnamefont {G.-L.}\ \bibnamefont
  {Ingold}},\ }\bibfield  {title} {\bibinfo {title} {Fundamental aspects of
  quantum brownian motion},\ }\href@noop {} {\bibfield  {journal} {\bibinfo
  {journal} {Chaos: An Interdisciplinary Journal of Nonlinear Science}\
  }\textbf {\bibinfo {volume} {15}},\ \bibinfo {pages} {026105} (\bibinfo
  {year} {2005})}\BibitemShut {NoStop}%
\bibitem [{\citenamefont {Vural}\ \emph {et~al.}(2018)\citenamefont {Vural},
  \citenamefont {Portalupi}, \citenamefont {Maisch}, \citenamefont {Kern},
  \citenamefont {Weber}, \citenamefont {Jetter}, \citenamefont {Wrachtrup},
  \citenamefont {L{\"o}w}, \citenamefont {Gerhardt},\ and\ \citenamefont
  {Michler}}]{vural2018two}%
  \BibitemOpen
  \bibfield  {author} {\bibinfo {author} {\bibfnamefont {H.}~\bibnamefont
  {Vural}}, \bibinfo {author} {\bibfnamefont {S.~L.}\ \bibnamefont
  {Portalupi}}, \bibinfo {author} {\bibfnamefont {J.}~\bibnamefont {Maisch}},
  \bibinfo {author} {\bibfnamefont {S.}~\bibnamefont {Kern}}, \bibinfo {author}
  {\bibfnamefont {J.~H.}\ \bibnamefont {Weber}}, \bibinfo {author}
  {\bibfnamefont {M.}~\bibnamefont {Jetter}}, \bibinfo {author} {\bibfnamefont
  {J.}~\bibnamefont {Wrachtrup}}, \bibinfo {author} {\bibfnamefont
  {R.}~\bibnamefont {L{\"o}w}}, \bibinfo {author} {\bibfnamefont
  {I.}~\bibnamefont {Gerhardt}},\ and\ \bibinfo {author} {\bibfnamefont
  {P.}~\bibnamefont {Michler}},\ }\bibfield  {title} {\bibinfo {title}
  {Two-photon interference in an atom--quantum dot hybrid system},\ }\href@noop
  {} {\bibfield  {journal} {\bibinfo  {journal} {Optica}\ }\textbf {\bibinfo
  {volume} {5}},\ \bibinfo {pages} {367} (\bibinfo {year} {2018})}\BibitemShut
  {NoStop}%
\bibitem [{\citenamefont {Kira}\ and\ \citenamefont
  {Koch}(2011)}]{kira2011semiconductor}%
  \BibitemOpen
  \bibfield  {author} {\bibinfo {author} {\bibfnamefont {M.}~\bibnamefont
  {Kira}}\ and\ \bibinfo {author} {\bibfnamefont {S.~W.}\ \bibnamefont
  {Koch}},\ }\href@noop {} {\emph {\bibinfo {title} {Semiconductor quantum
  optics}}}\ (\bibinfo  {publisher} {Cambridge University Press},\ \bibinfo
  {year} {2011})\BibitemShut {NoStop}%
\bibitem [{\citenamefont {Sinha}\ \emph {et~al.}(2019)\citenamefont {Sinha},
  \citenamefont {Meystre}, \citenamefont {Goldschmidt}, \citenamefont {Fatemi},
  \citenamefont {Rolston},\ and\ \citenamefont {Solano}}]{sinha2019non}%
  \BibitemOpen
  \bibfield  {author} {\bibinfo {author} {\bibfnamefont {K.}~\bibnamefont
  {Sinha}}, \bibinfo {author} {\bibfnamefont {P.}~\bibnamefont {Meystre}},
  \bibinfo {author} {\bibfnamefont {E.~A.}\ \bibnamefont {Goldschmidt}},
  \bibinfo {author} {\bibfnamefont {F.~K.}\ \bibnamefont {Fatemi}}, \bibinfo
  {author} {\bibfnamefont {S.~L.}\ \bibnamefont {Rolston}},\ and\ \bibinfo
  {author} {\bibfnamefont {P.}~\bibnamefont {Solano}},\ }\bibfield  {title}
  {\bibinfo {title} {Non-markovian collective emission from macroscopically
  separated emitters},\ }\href@noop {} {\bibfield  {journal} {\bibinfo
  {journal} {arXiv preprint arXiv:1907.12017}\ } (\bibinfo {year}
  {2019})}\BibitemShut {NoStop}%
\bibitem [{Note1()}]{Note1}%
  \BibitemOpen
  \bibinfo {note} {For details see the supplemental material.}\BibitemShut
  {Stop}%
\bibitem [{Note2()}]{Note2}%
  \BibitemOpen
  \bibinfo {note} {See supplemental material for the detailed
  derivation.}\BibitemShut {Stop}%
\bibitem [{\citenamefont {Bibbona}\ \emph {et~al.}(2008)\citenamefont
  {Bibbona}, \citenamefont {Panfilo},\ and\ \citenamefont
  {Tavella}}]{bibbona2008ornstein}%
  \BibitemOpen
  \bibfield  {author} {\bibinfo {author} {\bibfnamefont {E.}~\bibnamefont
  {Bibbona}}, \bibinfo {author} {\bibfnamefont {G.}~\bibnamefont {Panfilo}},\
  and\ \bibinfo {author} {\bibfnamefont {P.}~\bibnamefont {Tavella}},\
  }\bibfield  {title} {\bibinfo {title} {The ornstein--uhlenbeck process as a
  model of a low pass filtered white noise},\ }\href@noop {} {\bibfield
  {journal} {\bibinfo  {journal} {Metrologia}\ }\textbf {\bibinfo {volume}
  {45}},\ \bibinfo {pages} {S117} (\bibinfo {year} {2008})}\BibitemShut
  {NoStop}%
\bibitem [{\citenamefont {Reiter}\ \emph {et~al.}(2019)\citenamefont {Reiter},
  \citenamefont {Kuhn},\ and\ \citenamefont {Axt}}]{reiter2019distinctive}%
  \BibitemOpen
  \bibfield  {author} {\bibinfo {author} {\bibfnamefont {D.}~\bibnamefont
  {Reiter}}, \bibinfo {author} {\bibfnamefont {T.}~\bibnamefont {Kuhn}},\ and\
  \bibinfo {author} {\bibfnamefont {V.~M.}\ \bibnamefont {Axt}},\ }\bibfield
  {title} {\bibinfo {title} {Distinctive characteristics of carrier-phonon
  interactions in optically driven semiconductor quantum dots},\ }\href@noop {}
  {\bibfield  {journal} {\bibinfo  {journal} {Advances in Physics: X}\ }\textbf
  {\bibinfo {volume} {4}},\ \bibinfo {pages} {1655478} (\bibinfo {year}
  {2019})}\BibitemShut {NoStop}%
\bibitem [{\citenamefont {Carmele}\ and\ \citenamefont
  {Reitzenstein}(2019)}]{carmele2019non}%
  \BibitemOpen
  \bibfield  {author} {\bibinfo {author} {\bibfnamefont {A.}~\bibnamefont
  {Carmele}}\ and\ \bibinfo {author} {\bibfnamefont {S.}~\bibnamefont
  {Reitzenstein}},\ }\bibfield  {title} {\bibinfo {title} {Non-markovian
  features in semiconductor quantum optics: quantifying the role of phonons in
  experiment and theory},\ }\href@noop {} {\bibfield  {journal} {\bibinfo
  {journal} {Nanophotonics}\ }\textbf {\bibinfo {volume} {8}},\ \bibinfo
  {pages} {655} (\bibinfo {year} {2019})}\BibitemShut {NoStop}%
\bibitem [{\citenamefont {L{\"u}ker}\ and\ \citenamefont
  {Reiter}(2019)}]{luker2019review}%
  \BibitemOpen
  \bibfield  {author} {\bibinfo {author} {\bibfnamefont {S.}~\bibnamefont
  {L{\"u}ker}}\ and\ \bibinfo {author} {\bibfnamefont {D.~E.}\ \bibnamefont
  {Reiter}},\ }\bibfield  {title} {\bibinfo {title} {A review on optical
  excitation of semiconductor quantum dots under the influence of phonons},\
  }\href@noop {} {\bibfield  {journal} {\bibinfo  {journal} {Semiconductor
  Science and Technology}\ }\textbf {\bibinfo {volume} {34}},\ \bibinfo {pages}
  {063002} (\bibinfo {year} {2019})}\BibitemShut {NoStop}%
\bibitem [{\citenamefont {Loos}\ and\ \citenamefont
  {Klapp}(2019)}]{loos2019heat}%
  \BibitemOpen
  \bibfield  {author} {\bibinfo {author} {\bibfnamefont {S.~A.}\ \bibnamefont
  {Loos}}\ and\ \bibinfo {author} {\bibfnamefont {S.~H.}\ \bibnamefont
  {Klapp}},\ }\bibfield  {title} {\bibinfo {title} {Heat flow due to
  time-delayed feedback},\ }\href@noop {} {\bibfield  {journal} {\bibinfo
  {journal} {Scientific reports}\ }\textbf {\bibinfo {volume} {9}},\ \bibinfo
  {pages} {2491} (\bibinfo {year} {2019})}\BibitemShut {NoStop}%
\bibitem [{\citenamefont {Loos}\ \emph {et~al.}(2019)\citenamefont {Loos},
  \citenamefont {Hermann},\ and\ \citenamefont {Klapp}}]{loos2019non}%
  \BibitemOpen
  \bibfield  {author} {\bibinfo {author} {\bibfnamefont {S.~A.}\ \bibnamefont
  {Loos}}, \bibinfo {author} {\bibfnamefont {S.~M.}\ \bibnamefont {Hermann}},\
  and\ \bibinfo {author} {\bibfnamefont {S.~H.}\ \bibnamefont {Klapp}},\
  }\bibfield  {title} {\bibinfo {title} {Non-reciprocal hidden degrees of
  freedom: A unifying perspective on memory, feedback, and activity},\
  }\href@noop {} {\bibfield  {journal} {\bibinfo  {journal} {arXiv preprint
  arXiv:1910.08372}\ } (\bibinfo {year} {2019})}\BibitemShut {NoStop}%
\bibitem [{\citenamefont {Loos}\ and\ \citenamefont
  {Klapp}(2017)}]{loos2017force}%
  \BibitemOpen
  \bibfield  {author} {\bibinfo {author} {\bibfnamefont {S.~A.}\ \bibnamefont
  {Loos}}\ and\ \bibinfo {author} {\bibfnamefont {S.~H.}\ \bibnamefont
  {Klapp}},\ }\bibfield  {title} {\bibinfo {title} {Force-linearization closure
  for non-markovian langevin systems with time delay},\ }\href@noop {}
  {\bibfield  {journal} {\bibinfo  {journal} {Physical Review E}\ }\textbf
  {\bibinfo {volume} {96}},\ \bibinfo {pages} {012106} (\bibinfo {year}
  {2017})}\BibitemShut {NoStop}%
\bibitem [{\citenamefont {Calaj\'o}\ \emph {et~al.}(2019)\citenamefont
  {Calaj\'o}, \citenamefont {Fang}, \citenamefont {Baranger},\ and\
  \citenamefont {Ciccarello}}]{PhysRevLett.122.073601}%
  \BibitemOpen
  \bibfield  {author} {\bibinfo {author} {\bibfnamefont {G.}~\bibnamefont
  {Calaj\'o}}, \bibinfo {author} {\bibfnamefont {Y.-L.~L.}\ \bibnamefont
  {Fang}}, \bibinfo {author} {\bibfnamefont {H.~U.}\ \bibnamefont {Baranger}},\
  and\ \bibinfo {author} {\bibfnamefont {F.}~\bibnamefont {Ciccarello}},\
  }\bibfield  {title} {\bibinfo {title} {Exciting a bound state in the
  continuum through multiphoton scattering plus delayed quantum feedback},\
  }\href {https://doi.org/10.1103/PhysRevLett.122.073601} {\bibfield  {journal}
  {\bibinfo  {journal} {Phys. Rev. Lett.}\ }\textbf {\bibinfo {volume} {122}},\
  \bibinfo {pages} {073601} (\bibinfo {year} {2019})}\BibitemShut {NoStop}%
\end{thebibliography}
%

\end{document}